\documentclass[a4paper]{article}

\usepackage[english]{babel}
\usepackage[utf8x]{inputenc}
\usepackage[T1]{fontenc}

\usepackage[a4paper,top=3cm,bottom=2cm,left=3cm,right=3cm,marginparwidth=1.75cm]{geometry}

\usepackage{amsmath}
\usepackage{graphicx}
\usepackage[colorinlistoftodos]{todonotes}
\usepackage[colorlinks=true, allcolors=blue]{hyperref}
\usepackage{amssymb,amsfonts}
\usepackage{cite}
\usepackage{placeins}

\title{Osmotaxis in \textit{Escherichia coli} through changes in motor speed}
\author{Jerko Rosko$^1$,
Vincent Martinez$^2$, Wilson Poon$^2$ and 
Teuta Pilizota$^1$}

\begin{document}

\maketitle

$^1$ Centre for Synthetic and Systems Biology, Institute of Cell Biology, School of Biological Sciences, University of Edinburgh, Alexander Crum Brown Road, EH9 3FF, Edinburgh, UK

$^2$ Scottish Universities Physics Alliance and School of Physics and Astronomy, The University of Edinburgh, JCMB, Peter Guthrie Tait Road, Edinburgh EH9 3FD

\begin{abstract}
Bacterial motility, and in particular repulsion or attraction towards specific chemicals, has been a subject of investigation for over 100 years, resulting in detailed understanding of bacterial chemotaxis and the corresponding sensory network in many bacterial species. For \textit{Escherichia coli} most of the current understanding comes from the experiments with low levels of chemotactically-active ligands. However, chemotactically-inactive chemical species at concentrations found in the human gastrointestinal tract produce significant changes in \textit{E. coli's} osmotic pressure, and have been shown to lead to taxis. To understand how these nonspecific physical signals influence motility, we look at the response of individual bacterial flagellar motors under step-wise changes in  external osmolarity. We combine these measurements with a population swimming assay under the same conditions.  Unlike for chemotactic response, a long term increase in swimming/motor speeds is observed, and in the motor rotational bias, both of which scale with the osmotic shock magnitude. We discuss how the speed changes we observe can lead to steady state bacterial accumulation.
\end{abstract}

\section{Introduction}

Many bacterial species are not only able to self propel (exhibit motility), but also direct their motion towards more favorable environments. This behavior, called taxis \cite{Krell2011, Purcell1976}, has long been a subject of scientific investigation, as it serves a variety of purposes: seeking out nutrients and avoiding toxic substances \cite{WadArm2004, Adler1969}, identifying thermal \cite{PasterRyu2007} and oxygen \cite{AdlerM} gradients, as well as aiding pathogenic species in infecting their hosts \cite{RiveraChavez2013,Cullender2013}. The understanding of bacterial taxis is not only important when it comes to bacterial motility and accumulation; it also serves as a model for biological signal processing. Of particular interest are the precision \cite{Segall1986, Neumann2010}, sensitivity \cite{Cluzel2000} and robustness \cite{Yuan2012, Lele2012} that can be achieved with biological networks, and potentially utilized for human design purposes \cite{Navlakha2014, Babaoglu2006}. Specifically, bacterial chemotaxis, motion towards or away from specific chemicals \cite{WadArm2004}, was first described over 120 years ago \cite{Massart1889}. Since then, the systematic research efforts made it one of the best-studied topics in biology, especially when it comes to \textit{Escherichia coli}.

\textit{E. coli} swims by rotating a bundle of flagellar filaments \cite{Berg1973,TurnerBerg2000}, each powered by a bacterial flagellar motor (BFM), a rotary nano-machine that spins in the clockwise (CW) or counter-clockwise (CCW) direction \cite{SowaBerry2008}. Each bacterium possesses several individual motors randomly distributed along the cell body \cite{Tang1995,TurnerBerg2000}, which when rotating CCW enable formation of a stable filament bundle that propels the cell forward \cite{Berg2003}. When one or, most likely, a few motors switch to CW rotation, their respective filaments fall out of the bundle, leading to a tumble event \cite{TurnerBerg2000}. Forward swimming, likely in a different direction due to Brownian rotation of the bacterial cell during the tumble event, resumes when motors switch back to CCW direction and the bundle reforms \cite{Berg1973,TurnerBerg2000}. 

The probability of switching increases with the intracellular concentration of phosphorylated CheY protein (CheY-P) that interacts with the rotor of the BFM \cite{Welch1993}. CheY-P is a part of a feedback control circuit, the chemotactic network \cite{WadArm2004}, which relays outside information to the motor and allows \textit{E. coli} to direct its motion. Inputs of the circuit are the methyl-accepting chemotactic proteins (MCPs), transmembrane proteins that bind specific ligands in the cell exterior \cite{WadArm2004}, and through a signaling cascade affect the CheY-P to CheY ratio. When sensing attractants or repellents in $\mu$M range, the change in CheY-P to CheY ratio resets to the initial level within seconds, a characteristic feature of the network termed perfect adaptation \cite{Block1982}. Thus, directionality in the net motion of the cell arises through transient tuning of motor switching frequency in response to external stimuli \cite{WadArm2004}. 

The majority of work on \textit{E. coli} chemotaxis over the last 40 years has been performed in a minimal phosphate buffer (termed Motility Buffer \cite{Ryu2000}). However, one of the primary habitats of \textit{E. coli} is the gastrointestinal tract of humans and other warm-blooded animals \cite{Berg1996,Gordon2003}. This complex environment features not only various chemoattractants and repellents, but also spatial and temporal changes in osmolarity \cite{Fordtran1966, Datta2016,Begley2005}, which, in the stomach and small intestine of humans, reach up to 400 mOsmol. The exact composition and osmolarity depend on the meal, ingestion history and location within the gastrointestinal tract \cite{Fordtran1966}. 

Sudden osmotic increases, termed hyperosmotic shocks or upshocks, cause cell volume shrinkage and require solute pumping and/or synthesis to re-inflate the cell and re-establish osmotic pressure \cite{Wood2015, PilizotaShaevitz2012}. 
Non-specific spatial taxis away from sources of high concentrations, termed osmotaxis, has been observed in agar plates \cite{Li1988}. Osmotic stimuli can also send a signal down the network through mechanical stimulation of chemoreceptors  \cite{Vaknin2006}; yet, osmotaxis was observed in gutted mutants lacking all chemotactic network components \cite{Li1993}.

To clarify the exact nature of osmotactic response, we study the phenomenon on both single cell and population levels, observing the rotation of individual flagellar motors under stepwise increases in osmolarity, and measuring the swimming speeds of a population of $\sim$ 10000 bacteria after exposure to an osmotic shock. The shock magnitudes we administer mimic those encountered in the human gastrointestinal tract \cite{Fordtran1966}. We find that a stepwise increase in osmolarity results in an elevated CCW-CW switching frequency that scales with the shock magnitude. In addition, we observe osmokinesis post osmotic shock,~i.e. significant changes in the motor, and consequently, swimming speeds of bacteria. Lastly, for higher shock magnitudes, we observe a loss of motor speed immediately post shock, followed by a transient, attractant-like response that is coupled with a speed recovery.
We discuss how the observed non-adaptive response and osmokinesis can lead to taxis.


\section{Results}
\subsection{Single motor response to an osmotic shock is complex}
We begin with the reasonable assumption that at least one component of the chemotactic network responds to an osmotic stimulus in order to generate previously observed osmotactic behavior \cite{Li1988}. Then, the response should be evident in the output of the chemotactic network, the Clock-Wise (CW) bias of a single BFM. CW bias is defined as the fraction of time the BFM spends rotating in the clockwise direction \cite{Bai2010}:
\begin{equation}\label{eq:bias}
CW Bias=\frac{N_{cw}}{N_{tot}}
\end{equation}
where $N_{cw}$ is the number of data points corresponding to CW rotation and $N_{tot}$ is the total number of data points in a given time interval.

To compute the \textit{CW Bias}, we measured the speed of an individual BFM when exposed to a step-wise increase in the external osmolarity. SI Appendix Fig.~5 gives a schematic of the bead assay used for measuring the motor speed \cite{Ryu2000, Bai2010}. Briefly, we attach cells to the cover slip and attach a latex bead, 0.5 $\mu$m in diameter, to a short filament stub \cite{Ryu2000, Bai2010}. The rotation of the BFM-driven bead is recorded using back-focal-plane interferometry at a 10~kHz sampling rate \cite{Pilizota2007,Denk1990,Svoboda1993}. A representative single-motor rotation trace so obtained is shown in Fig.~1A. The shaded light blue interval shows the motor speed and rotational direction prior to osmotic stimulus. Positive motor speed values represent CCW rotation and negative CW. Fig.~1B shows the histogram of the \textit{CW Bias} obtained from the motor speed trace in A (\textit{Methods}). Prior to the stimulus, the motor switches from CCW to CW rotation $\sim$ 4 times per minute at $CW Bias=0.07$, agreeing with previous studies \cite{Bai2010, Bai2013}.

At $\rm t=5 min$, the extracellular osmolarity is elevated by delivering 400~mM sucrose, at a local flow rate of $\rm0.68\ \mu l/min$ \cite{Buda2016}, causing an osmotic upshock of 488~mOsmol/kg. Immediately upon upshock, the motor stops switching rotational direction and speed drops. Then, a recovery phase begins (Fig.~1A and B, non-shaded time interval) with the motor speed gradually increasing while switching is absent. Post-recovery phase, which we define as the period after motor switching has resumed, is characterized by an increased switching frequency. Consequentially, the \textit{CW Bias} is also elevated (to 0.16), and is maintained over at least $~20$~min, suggesting that, unlike chemotaxis, the osmotactic response does not exhibit perfect adaptation \cite{Segall1986,Block1982}. The color map at the bottom of Fig.~1B is a compact representation of the histogram in Fig.~1B. 

To determine if CheY-P is necessary for the osmotic response observed in Fig.~1A and B, we performed single-motor measurements on a strain lacking CheY ($\rm\Delta CheY$ mutant \cite{Turner1993, Fahrner1995}). A representative trace is shown in Fig.~1C. The Recovery Phase of the $\Delta CheY$ mutant is the same as observed in Fig.~1A for the chemotactic wild type. However, the Post-recovery Phase of the mutant shows no switching events, indicating that CheY-P is necessary for the elevated bias observed in Fig.~1A and B.

\begin{figure}[p]
\centerline{\includegraphics[width=1.0\linewidth]{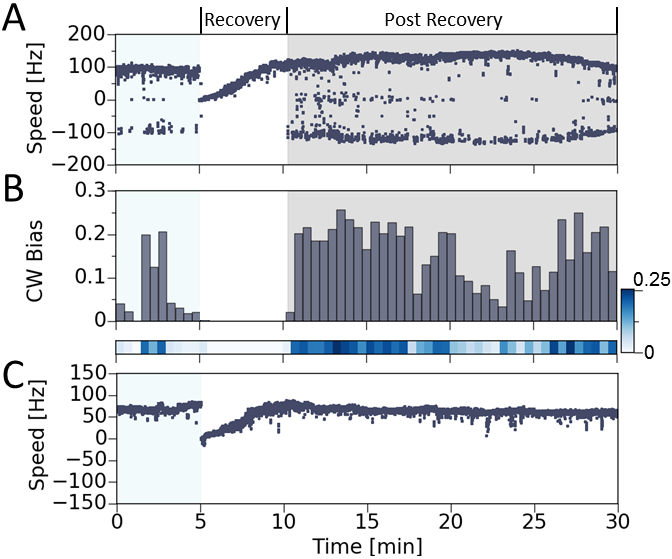}}
\caption{(A) An example 30~min speed trace obtained from a single BFM.  The cell was initially in VR Buffer (\textit{Methods}), indicated in shaded light blue, and exposed to an osmotic upshock of 488~mOsmol/kg at t=5~min.  The shock was delivered as a step increase by flowing in VRB containing additional 400~mM Sucrose, resulting in shock magnitude of 488~mOsmol/kg. (B) A histogram of clockwise bias computed by binning the trace in Fig.~1A into 30 second bins and dividing the time spent rotating clockwise by the bin length (see also equation \ref{eq:bias} and \textit{Methods}). Below is the same histogram condensed into a color map, with an intensity scale to the right. (C) Single-motor speed trace of a $\Delta CheY$ mutant exposed to the same osmotic upshock as in Fig.~1A.}\label{Figure1}
\end{figure}

\subsection{Osmotactic response does not exhibit perfect adaptation}
We analyzed 69 cells exposed to three up-shock magnitudes to confirm the absence of perfect adaptation observed in Post-recovery Phase (Fig.~1A and B). Up-shocks were delivered as in Fig.~1, by exchanging VR Buffer (\textit{Methods}) with the same buffer supplemented with 100, 200 and 400~mM sucrose, corresponding to osmotic shocks of 111, 230 and 488~mOsmol/kg (similar to osmolalities found in the small intestine \cite{Fordtran1966}, see also SI Appendix for osmolality measurements of all our buffers). 

Fig.~2A shows color map histograms of each individual single-motor bias trace, where darker color represents higher \textit{CW Bias} and white represents smooth swimming. The Recovery Phase, characterized by motor speed recovery and zero \textit{CW Bias} period, scales with the shock magnitude. SI Appendix Fig.~6~A shows the duration of the Recovery Phase, $T_{rec, CW}$, against the shock magnitude, where for the highest shock administered $T_{rec, CW} \sim$5~min.  The length of the Recovery Phase roughly corresponds to the time \textit{E. coli} takes to recover its volume and osmotic pressure upon the hyperosmotic shock \cite{Pilizota2014}. 
Throughout the Post-Recovery Phase, \textit{CW Bias} levels do not, on average, relax to their initial pre-shock values. In our measurements this phase lasts for $\sim$ 10-20~min, which is significantly longer than chemotactic adaptation times \cite{Segall1986} even when saturating attractant concentrations are used \cite{Berg1975}. 

Evidence for this assertion is shown in the \textit{CW Bias} histograms in Fig.~2B, which shows a distribution computed from 5~min recordings in VR Buffer prior to shock, and Fig.~2C, which shows a distribution computed from 3~min intervals taken at various time points, 12 or more minutes after shock with 200~mM sucrose. Total of 120 single-motor recordings were used for the pre-shock condition and 96 for after. The median value of the population \textit{CW Bias} shifts from 0.01 pre-shock to 0.06 after addition of 200~mM sucrose. 
Fig.~2D shows the median \textit{CW Bias} in time for each of the three different shocks, calculated from the raw speeds of individual cells presented in Fig.~2A, using a 60~s wide moving window. Here, only medians are shown for clarity and the means, together with the interquartile range for each shock magnitude, are plotted in SI Appendix Fig. 6~B-D. While the \textit{CW Bias} shows some recovery in time, in particular for 111~mOsmol/kg and 230~mOsmol/kg shocks, it proceeds on a slow time scale and the median values at the end of our measurement time remain elevated with respect to the initial value.

A further corroboration of long-term increase in \textit{CW Bias} post osmotic shock comes from separating the \textit{CW Bias} values of Fig. 2B according to the time point at which they were measured, relative to the administration of the osmotic shock (t=0). This is displayed in the inset of Fig. 2B and shows that the elevated \textit{CW Bias} persists over a time scale as long as 1~h.

\begin{figure*}[p]
\centerline{\includegraphics[width=1\linewidth]{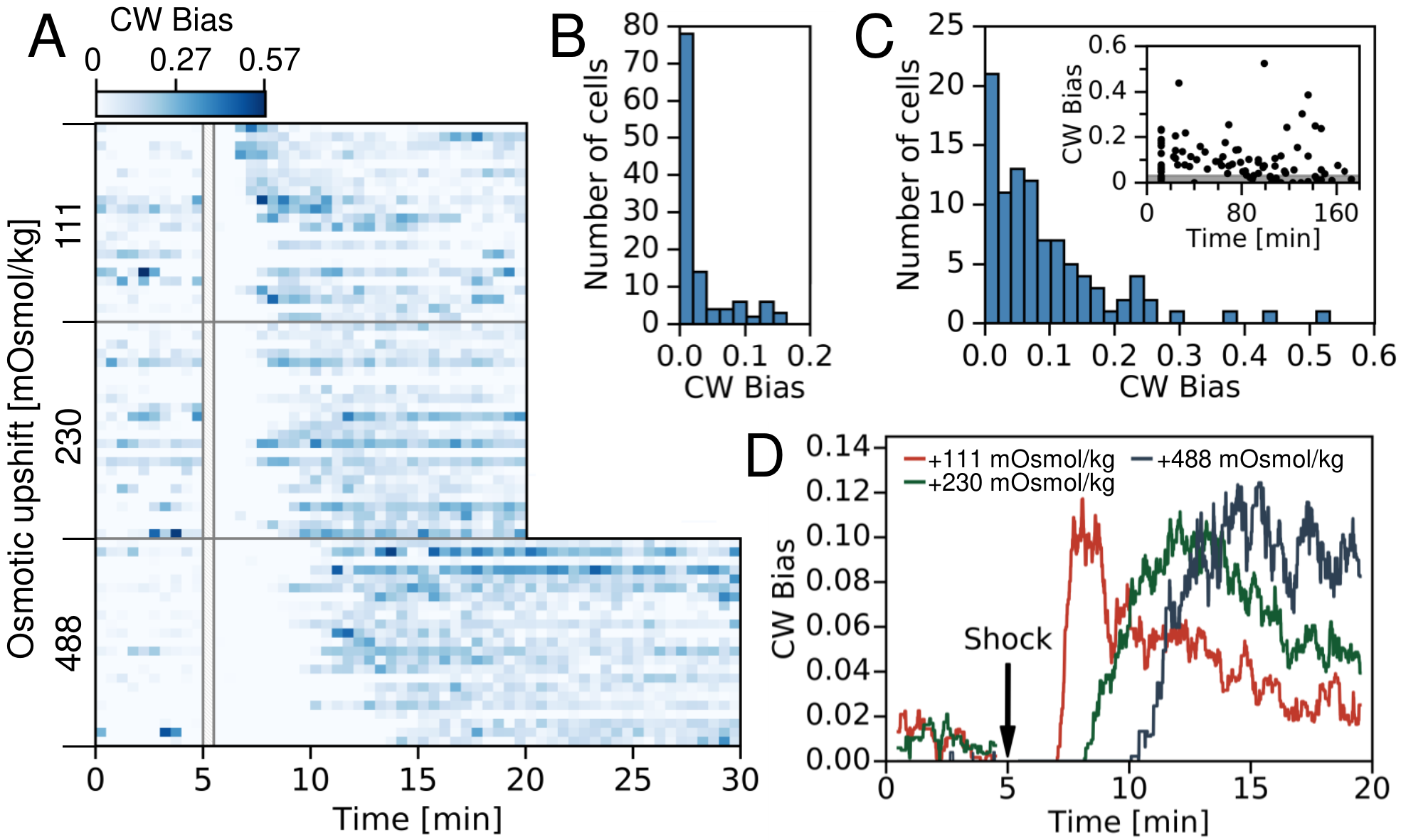}}
\caption{(A) Stacked single cell color maps (histograms) of \textit{CW Bias} for three different shock magnitudes (indicated on the left). Bin widths are 30~s. VR Buffer was exchanged for the same buffer with the addition of sucrose at t=5~min. The white hatched column represents the period of the media exchanged whose duration was $\sim$10-15~s. 22, 24 and 23 cells are given for the 111, 230 and 488~mOsmol/kg condition, respectively. Color map scale is given at the top. (B) Histogram of \textit{CW Bias} for cells prior to osmotic upshock (in VR Buffer) and (C) post osmotic upshock, administered by exchanging VRB with VRB + 200~mM sucrose. Bin width is set to 0.020. Total of 120 motors (each on a different cell) were used to construct Fig.~2B and 96 motors for Fig.~2C. Inset in Fig.~2C plots these 96 single motor biases against the time after their respective osmotic upshift. Gray shading represents the range between 25$\rm ^{th}$ and 75$\rm ^{th}$ percentile of the CW bias distribution in VR Buffer given in B. (D) Median population \textit{CW Bias} in time, computed from cells given in Fig.~2A, for different shock magnitudes: 111~mOsmol/kg (red), 230~mOsmol/kg (green) and 488~mOsmol/kg (dark blue). Black arrow indicates the time at which hyperosmotic shock was administered.} \label{Figure2}
\end{figure*}

\subsection{Osmotic response shows osmokinesis, i.e. changes in motor speed}
The response to an osmotic upshock includes not only \textit{CW Bias} dynamics, but also changes in motor and free-swimming speed. Fig.~3A shows a color map plot of normalized single-motor speeds of the same 69 cells presented in Fig.~2A. All cells were originally in VR Buffer and subsequently exposed to an osmotic upshock using 100, 200 or 400~mM sucrose, corresponding to upshifts of 111, 230 and 488~mOsmol/kg. BFM speeds were normalized with respect to the initial speed of each motor, i.e. to the average value of the first 15 seconds of the recording.

Following the osmotic upshock, motors show two kinds of behavior. If a shock is of a large magnitude, such as 488~mOsmol/kg, the speed drops sharply and significantly, with a phase of speed recovery that follows. As can be seen from Fig.~3A and B the speed recovery ($T_{rec,\omega}$) lasts $\sim$4.4~min on average (5.2~min median). Additionally, this is of similar magnitude as $T_{rec,CW}$ (5.4 ~min mean, 5.0~min median, SI Appendix Fig.~6) and likely corresponds to the period of post-hyperosmotic shock volume recovery \cite{Pilizota2014}. After recovery, the speed increase continues, leading to elevated levels compared to the pre-shock values, Fig.~3C. 
Weaker upshocks, 111 and 230~mOsmol/kg, are characterized by an increase in motor speed without a significant speed drop, as seen in Fig.~3A-C. The 0~mOsmol/kg condition is a buffer to buffer control flush, where we used the same shocking protocol as for osmotic upshocks.

\begin{figure*}[p]
\centerline{\includegraphics[width=1\linewidth]{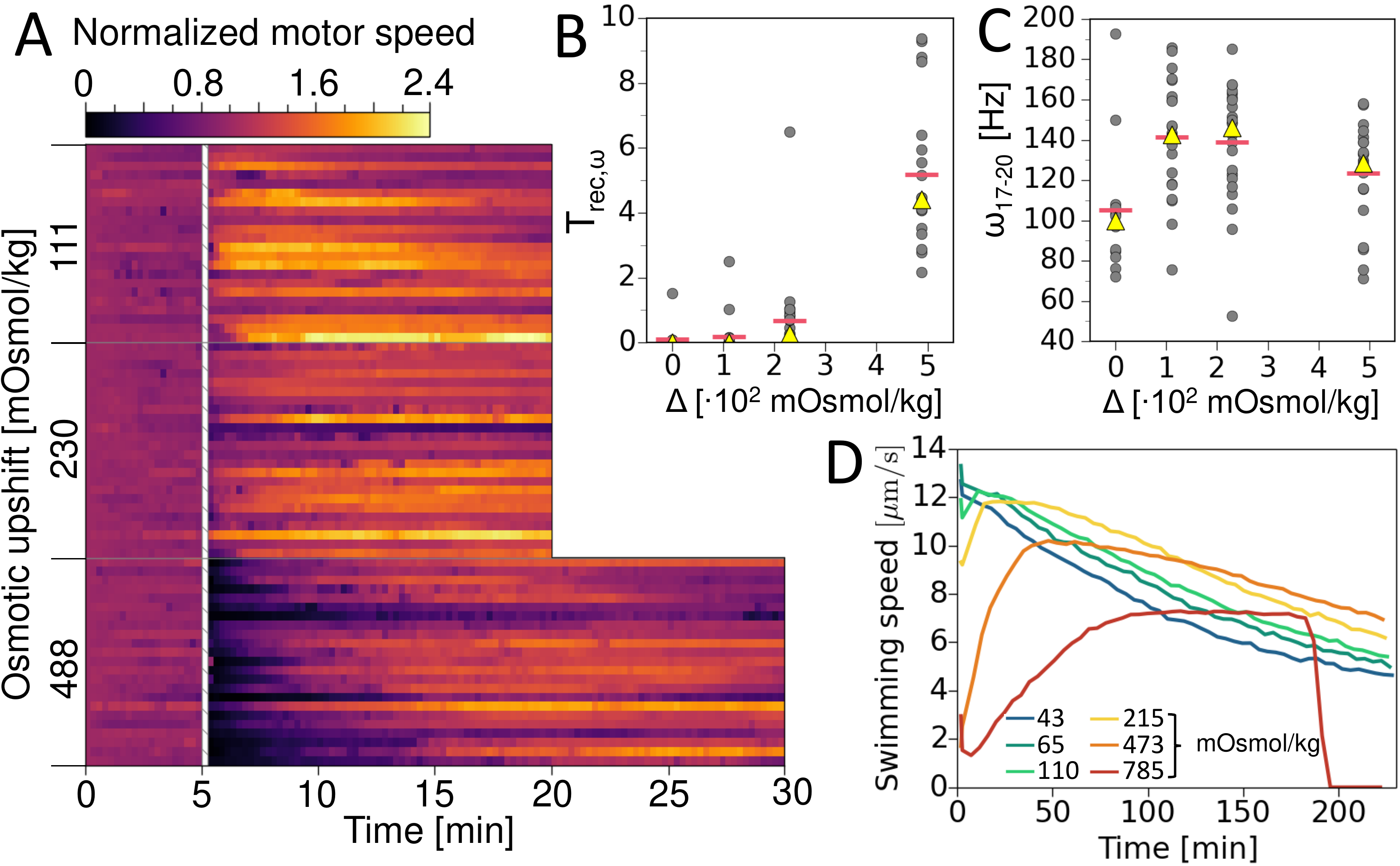}}
\caption{(A) Stacked single-motor (also single-cell) speed histograms, where the motor speed for each BFM is normalized to the average value of the first 15~s. Bin widths are 15~s and the color represents the bin height. Results are grouped by upshock magnitude, as indicated on the left hand edge. The white hatched column represents the point where an osmotic shock was performed by exchanging VR Buffer for VR Buffer + sucrose and the flow lasted for 10-15~s. 22, 24 and 23 cells are given for the 111, 230 and 488~mOsmol/kg conditions, respectively. The color map scale is given at the top of the figure. (B) Time necessary to recover the average value of pre-shock speed. Red horizontal bars are mean, and yellow triangles are median values, and the graph contains 18, 22, 23 and 20 single motor data points for the 0, 111, 230 and 488~mOsmol/kg upshocks. One value for the 230~mOsmol/kg condition and three for the 488~mOsmol/kg have been excluded from the graph as these motors do not recover average initial speed in the course of recording. The 0~mOsmol/kg condition is a buffer to buffer control flush. (C) Single motor speeds calculated as 3~min averages corresponding to a section between t=17 and t=20~min in A. The graph contains 12, 22, 24 and 23 single motor data points for the 0, 111, 230 and 488~mOsmol/kg upshocks. The 0~mOsmol/kg condition contains 12 out of 18 control flushes that were at least 20~min long. (D) DDM measurement of swimming speeds following an osmotic shock. Cells were shocked in microfuge tubes and brought into a microscope within 2 minutes. The legend shows shock magnitudes and the mean speed is the average of swimming speeds obtained for each time point in a range of different length scales (\textit{Methods}). The systematic error of our measurements is then calculated as the standard deviation of the mean, and falls within $\sim$ 5\% of the mean value. Here it was not plotted for clarity.
}\label{Figure3}
\end{figure*}

To explore the population-level significance of our observation of osmokinesis in single cells, we next performed Differential Dynamic Microscopy (DDM) (see \textit{Methods}) \cite{Wilson2011,Martinez2012}. DDM is a fast, high-throughput method for measuring the distribution of swimming speeds in populations of a range of different self-propelled particles, averaging over up to $\sim 10^4$ particles at the same time. The technique is well suited for rapid scanning of parameter space, so that it also allowed us to extend the range of external osmolarities studied. DDM characterizes the motility of
a population of particles (in our case \textit{E. coli}) by analyzing the statistics of temporal fluctuations of pixel intensities in a sequence of low-optical resolution microscopy images, where the intensity fluctuations are caused by the variation in number density of particles. Specifically, we measure the differential image correlation function (DICF), which is effectively a power spectrum of the difference between two images taken at separate time points \cite{Wilson2011,Martinez2012}. If a theoretical motility model exists, such as in the case of \textit{E. coli} \cite{Berg2003}, the expected DICF can be calculated and fitted to experimental data \cite{Wilson2011}, allowing accurate estimates of the distribution of free-swimming speeds of a large number of bacteria \cite{Martinez2012}. 

Prior to DDM measurements cells were kept in VRB and osmotically shocked in a microfuge tube. Upon the upshock cells were quickly placed into a capillary for DDM measurements and the capillary was sealed, resulting in a fixed amount of oxygen present during the experiment. Swimming speed recordings commenced within 2~min after the upshock (\textit{Methods}) and are shown in Fig.~3D. The gradual decrease of swimming speed with time observed in Fig.~3D for all magnitudes of osmotic upshocks was previously characterized in Motility Buffer, where \textit{E. coli} maintains Proton Motive Force (PMF) using endogenous energy sources \cite{Dawes1965,SwarzLinek2016}. At fixed buffer composition, the time it takes to consume all available oxygen is inversely proportional to the cell concentration, and upon oxygen exhaustion a sudden `crash' in swimming speed occurs \cite{SwarzLinek2016}. Interestingly, we do not see such a `crash' in swimming speed for the lowest five values of the imposed upshock, but do see a `crash' when the upshock is at the highest value of 785~mOsmol/kg. Since the cell concentration is fixed, this implies that cells consume oxygen at a significantly higher rate at the highest osmotic shock.

The increase in BFM speed observed in single cells, Fig.~3A, translates to an increase in population swimming speed with increasing upshock magnitudes, Fig.~3D. Similarly, in agreement with Fig.~3A, for high shock magnitudes, in particular for 473 mOsm/kg and 785 mOsm/kg sucrose upshocks, a sharp speed drop is observed immediately upon upshock. We also see a Recovery Phase, with duration increasing with the shock magnitude. 

Increasing sucrose concentrations results in the increased viscosity of the media, consequently increasing the drag coefficient on the motor (measured values of viscosity of our solutions are given in SI Appendix). In Fig.~3 we show BFM speeds and cell swimming speeds uncorrected for this effect, as the actual speed of the cell will be relevant for taxis and effective diffusivity. The viscosity corrected BFM speeds, calculated under the assumption that the motor torque does not change with the increasing viscosity, are shown in SI Appendix Fig.~8. 

Furthermore, we check for the presence of steps during single motor speed recovery. To that end, in SI Appendix Fig.~9 we show examples of BFM speeds during the Recovery Phase after 488~mOsmol/kg upshock, starting just after the osmotic shock was administered. Majority of the traces do not show obvious steps during the BFM speed Recovery Phase.

\section{Discussion}
\subsection{Origins of osmotaxis}
With no external stimuli \textit{E.coli} swims in an almost straight line and re-orients every so often in a nearly-random fashion, performing a random walk with an effective diffusion constant $D\sim v^{2}/\alpha$ \cite{Schnitzer1993, Tailleur2008}, where $v$ is the swimming speed and $\alpha$ the tumbling rate given by

\begin{equation}\label{eq:deff2}
\alpha=\frac{3(1-\cos{\phi_0})(\tau_{\rm run}+\tau_t)}{\tau_{\rm run}^2}
\end{equation}

with $\tau_{run}$ mean run time, $\phi_0 \backsim 71^{\circ}$ mean reorientation angle following a tumble, and $\tau_{t}$ the duration of the tumble event \cite{Schnitzer1993,Lovely1975}.

Upon sensing a sudden increase in attractant concentration, \textit{E. coli} sharply elevates the CCW bias for $\sim$ 1~s and then slightly lowers it for $\sim$ 3~s, returning to the prestimulus level in $\sim$ 4~s \cite{Block1982}. It is this characteristic impulse response of modulating the tumbling rate that allows \textit{E. coli} to navigate towards a favorable environment. The detailed chemotactic performance and its origins have often been studied \cite{Schnitzer1993, Clark2005, deGennes2004, Strong1998, Cates2012, Schnitzer1990, Celani2010}. 

The observation that a chemotactic mutant of \textit{E. coli} (\textit{CheRCheB}) lacking normal methylation and demetylation enzymes does not show chemotactic accumulation, but does respond to raised attractant concentrations by lowering the tumble rate without adaptation \cite{Block1982,Segall1986} lead researchers to calculate the consequences of spatially varying $\alpha(x)$ and $v(x)$ on steady state bacterial density \cite{Schnitzer1993,Schnitzer1990}. Similar calculations were performed for particles with $v$ and $\alpha$ dependent on the local density \cite{Tailleur2008}, or when studying diffusion of bacteria in porous media \cite{Licata2016}. These calculations offer possible explanations for previously observed osmotaxis, given the characteristic speed and bias response we here observed.
 
If we assume the simplest scenario, where there is no directional dependency of $\alpha (x)$ and $v(x)$; the steady state \textit{E. coli} density will be inversely proportional to its swimming speed \cite{Tailleur2008}:

\begin{equation}\label{eq:rhov}
\rho (x) \sim \frac{1}{v(x)}
\end{equation}

\begin{figure}[p]
\centerline{\includegraphics[width=1.0\linewidth]{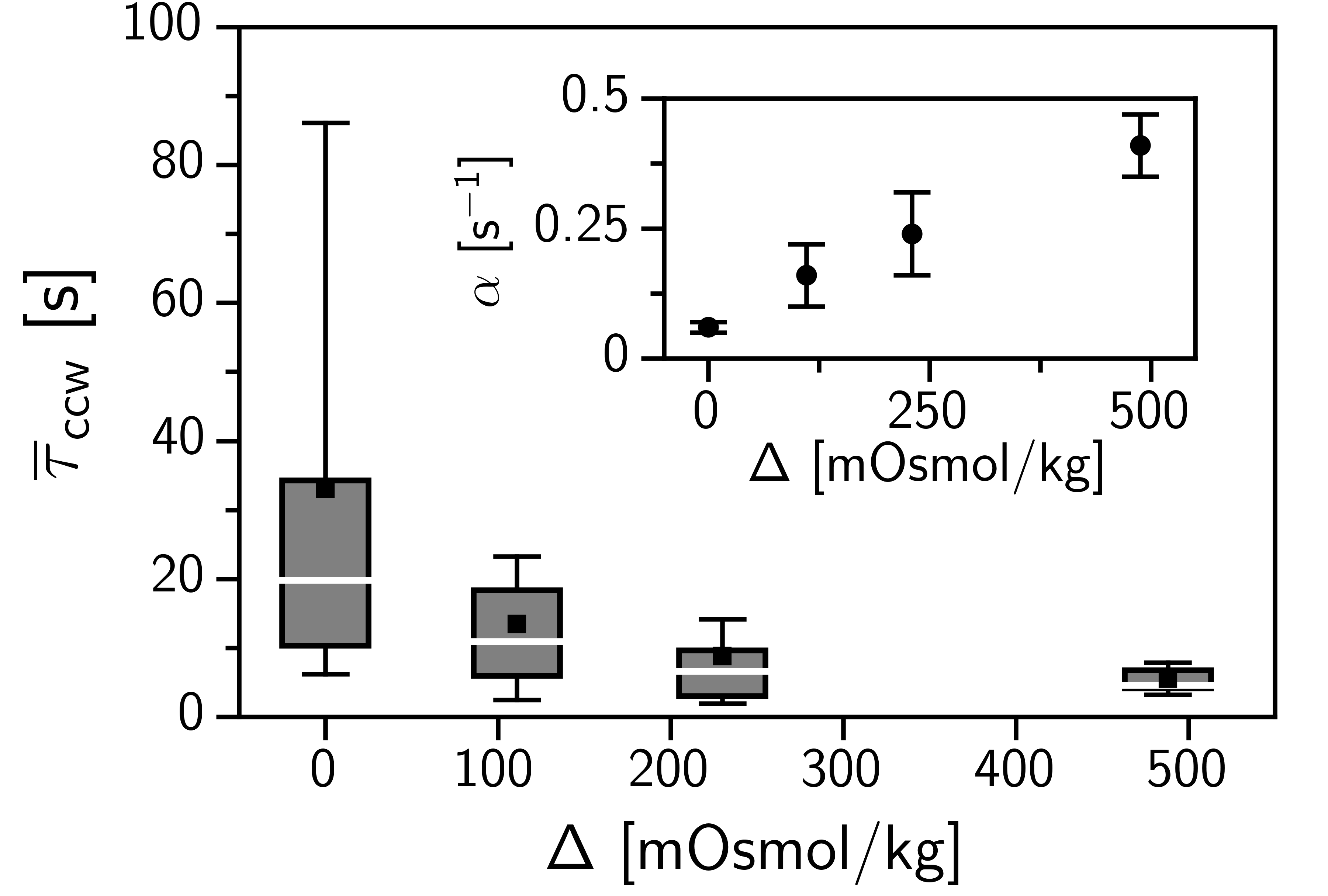}}
\caption{Mean counter clockwise motor interval, corresponding to cell runs, as a function of osmotic upshift. The "0" condition box contains 96 mean run intervals, calculated by averaging interval lengths over 5~min before osmotic upshift. Subsequent conditions, osmotic upshifts of 111, 230 and 488~mOsmol/kg, contain 21, 20 and 22 single-motor values calculated by averaging interval lengths over the time span from t=12~min to t=15~min after an osmotic shock. Not all cells were used to obtain $\overline{\tau}_{CCW}$ as note all cells had bound CCW intervals (see SI Appendix Fig.~11).  In the inset, we calculate the tumbling rate $\alpha$ according to equation \ref{eq:deff2}, approximating the run times with $\bar{\tau}_{ccw}$ and tumble times with $\bar{\tau}_{cw}$ given in SI Appendix Fig.~10.}\label{Figure4}
\end{figure}

In Fig.~3C we show motor speeds, $\omega$, and in Fig.~4 mean CCW motor interval ($\overline{\tau}_{CCW}$) and tumbling rate ($\alpha$) at a given osmolality. The $\omega$ and $\overline{\tau}_{CCW}$ were obtained from the last 3~min of the post-osmotic shock recordings presented in Fig. 3A, which represent the long term osmolarity dependent changes.  We then calculated $\alpha$ from $\overline{\tau}_{CCW}$ and $\overline{\tau}_{CW}$ (Fig.~4 and SI Appendix Fig.~10) using equation \ref{eq:deff2}. The assumption was made that swimming speed $v$ and the mean run time $\overline{\tau}_{run}$ are proportional to the motor speed $\omega$ and the mean CCW interval $\overline{\tau}_{CCW}$, in agreement with previous studies \cite{Magariyama1995}, and our own DDM and motor measurements given in Fig.~3. 

Thus, based on experimental results we present here and previous theoretical calculations (equation \ref{eq:rhov} \cite{Tailleur2008,Schnitzer1993}) we would expect accumulation at lower osmolarities, equivalent to negative taxis previously observed \cite{Li1988}. The changes in tumbling rates suggest differences in the time dependent approach to steady state, but not the steady state density distribution itself \cite{Schnitzer1993, Schnitzer1990}.  

The theoretical calculations we refer to make the assumption that spatial gradients of $v(x)$ and $\alpha (x)$ are small \cite{Tailleur2008,Schnitzer1993}, such that we do not expect local gradients in steady state bacterial densities. For steeper osmotic gradients we expect a more complex osmotactic response, in particular since the Recovery Phases shown in Fig.~2 and Fig.~3 will no longer be vanishingly short, as we assumed above, which could explain some previous contradictory osmotactic observations \cite{Li1993}.

Our assumption that $\alpha (x)$ and $v(x)$ lack directional dependence in the case of osmotaxis, could be satisfied if the observed changes in $\alpha$ are not due to signaling within the chemotactic network, but purely due to changes in the interaction of CheY-P with the rotor units of the motor. For example, previous observations indicate that at higher external osmolarities, osmotic pressure is kept the same \cite{Pilizota2014}. Thus, the crowding in an already crowded cytoplasm \cite{Parry2014} increases, which can affect cytoplasmic interactions \cite{Klump2013,Paudel2014} and the binding of CheY-P to the BFM, perhaps in a similar fashion to that observed at higher hydrostatic pressures \cite{Nishiyama2013}.

However, it is also possible that the observed changes in  $\overline{\tau}_{CCW}$ and $\omega$ are a consequence of signal processing by the chemotactic network. In particular, Vaknin \textit{et al.} reported that osmotically induced changes in cell volume can perturb the chemoreceptors, and that the signal could travel down some of the network components \cite{Vaknin2006}. In such a scenario, $\alpha (x)$ and $v(x)$ would be directionally dependent (as in chemotactic response) and steady-state cell density would need to be calculated taking into account the directional dependency as well as the characteristic signal response we observed. Future work needs to investigate strains lacking specific parts of the network to determine the contribution of signaling and different components of the network to the osmotactic response.

\subsection{Origins of osmokinesis}
The large increase in motor speed post osmotic shock could be due to increases in PMF, possibly through alterations in cell metabolism; or, as an alternative explanation, due to the increase in number of stator units through mechanosensation \cite{Lele2013, Tipping2013} or adaptive motor remodeling \cite{Yuan2012}. The BFM has been shown to act as a mechanosensor, increasing the number of stator units in response to higher loads \cite{Lele2013, Tipping2013}. As the viscosity of the medium rises with addition of sucrose, the speed increase we observed could be due to the incorporation of additional stator units. In fact, at higher loads, an increase in the mean CW interval has been reported as well \cite{Fahrner2003, Lele2013}. At our load, a 0.5~$\mu m$ bead attached to the motor via a short filament stub, the motor is still expected to operate in the high-load, 'plateau' region of the torque-speed curve \cite{Inoue2008, LoThesis} with the estimated full stator number \cite{LoThesis}. In addition, we observe an average 30 Hz increase in motor-speeds even at our lowest osmotic shocks where the viscosity of the solution hardly changes (it increases by 1.057 times) and even free-swimming cells with the motor operating in the linear-torque regime do not increase the swimming speed at viscosity we use in our experiments \cite{Martinez2014}. Therefore, it is less likely that additional stator incorporation or adaptive motor remodeling are sole explanations for the speed increases we observe.

SI Appendix Fig.~9 shows the motor speeds during the speed recovery phase for all of the BFMs recorded in the +488~mOsmol/kg condition. In majority of the traces no obvious steps are observed, suggesting that the increase in motor speed could be due to the increase in PMF with full set of stators present. Here we note that stator engagement with the rotor has been reported as torque dependent \cite{Tipping2013}, where in absence of torque (motor rotation) stators disengage from the rotor. We would then expect stator resurrection during the Recovery Phase shown in SI Appendix Fig.~9. Absence of obvious step-wise increases indicates that volume shrinkage caused by the osmotic shock could affect motor dynamics and perhaps prevent, or slow down, stator disengagement.

The length of the motor Recovery Phase observed in Fig.~2 and 3 is $\sim$5-15~min, in line with the volume recovery timescales observed previously post osmotic upshocks \cite{PilizotaShaevitz2012, Pilizota2014}. The timescales of speed recovery (at similar shock magnitudes) observed from DDM data in Fig.~3D are longer, with recovery lasting $\sim$40~min for the 473~mOsmol/kg upshock. Some variation in these times scales can be due to the difference between individual motor speed recovery and subsequent bundle formation. However, based on Fig.~3D, we suspect that greater contribution to the difference comes from alterations in oxygen consumption rate, which in turn shifts the time it takes to reach the maximum swimming speeds.
\newline

By looking at individual BFMs and population swimming speeds together, we reveal the main characteristics of \textit{E. coli's} motility response to step increases in external osmolarity. The response consists of long term \textit{CW Bias} and motor rotation/cell swimming speed increase. This is the first observation of chemokinesis (osmokinesis) in \textit{E. coli}. We discuss how such observed speed increases can lead to negative taxis previously reported. Our study emphasizes the importance of investigating bacterial motility in environments that mimic natural habitats, in the effort to understand the role and evolutionary advantage swimming offers to bacterial cells \cite{Lackraj2016,Gauger2007,Tamar2016}.



\section{materials}
\subsection{\textit{E. coli} strains and plasmids}
\textit{E. coli} strains KAF84 and KAF95 \cite{Turner1993} were used for BFM speed and bias, and MG1655 \cite{Blattner1997} for DDM experiments. Both KAF84 and KAF 95 carry the \textit{fliC726} allele, (produce nonflagellate phenotypes), and contain a plasmid carrying an ampicillin resistance and a \textit{fliC}$^{sticky}$ gene (produces flagellar fillaments that stick readily to surfaces). Additionally, KAF84 is a chemotactic wild type and KAF95 is a $\Delta$CheY strain and therefore can not perform chemotaxis, producing a smooth swimmer phenotype. MG1655 is a K-12 strain and a chemotactic wild type.

\subsection{\textit{E. coli} Growth and culturing}
KAF95, KAF84 and MG1655 cells were grown in Tryptone Broth (1\% Bacto tryptone, 0.5\% NaCl) at 30$^{\circ}$C while shaken at 200 RPM \cite{Bai2010,Martinez2012}. KAF95 and KAF84 were supplemented with 100 $\mu$g/ml of ampicillin and grown to OD=0.8-1.0 (Spectronic 200E Spectrophotometer, Thermo Scientific, USA), and MG1655 to OD=0.6. 
After growth cells were washed into VR buffer (\textbf{V}olume \textbf{R}ecovery) composed of the Modified Motility Buffer (MMB), which is 10~mM sodium phosphate buffer, pH=7.1 (an aqueous solution with 6.1mM of Na$_2$HPO$_4$, 3.9 mM of NaH$_2$PO$_4$) and 0.01~mM of Ethylenediaminetetraacetic acid (EDTA), with added Glycine Betaine, Potassium Chloride and Choline Chloride to final concentrations of 10, 20 and 10~mM, respectively. These compounds act as osmoprotectants and allow the cell to recover volume after an osmotic shock. MMB is a variant of the Motility Buffer, commonly used in flagellar motor and chemotaxis experiments \cite{Ryu2000, Bai2010}, with sodium phosphates substituted for potassium phosphates. After washing, all the experiments were performed in VR buffer. KAF95 and KAF84 cells were washed by centrifuging them into a pellet and exchanging solution, while MG1655 cells were washed by gentle filtration to preserve filaments \cite{SwarzLinek2016} and experiments were performed in VR buffer as for motor speed measurements.

\subsection{Sample preparation and osmotic shock} For BFM experiments flagellar filaments were truncated by passing a bacterial suspension through two syringes with narrow gauge needles (26G) connected with a plastic tube ('shearing device' \cite{Bai2010, Ryu2000,Pilizota2009}). Subsequently, cells with truncated filaments were washed by centrifugation. Slides for BFM experiments were prepared as before \cite{Bai2010, Pilizota2009} by layering two parallel strips of double sided sticky tape onto a microscope slide and covering them with a cover glass, forming a tunnel slide (SI Appendix Fig.~5) of approximate volume of $\sim$ 8~$\mu$l. 1\% poly-L-lysine was loaded into the tunnel and extensively washed out after keeping it in for $\sim$10~s to allow glass coating. Cells with truncated filaments were loaded into the tunnel and incubated for 10~min in a humid environment to prevent evaporation. Subsequently, non attached cells were washed out. Next, 0.5~$\mu$m beads in diameter (Polysciences) were added and incubated for 10~min to allow sticking to the filaments and excess beads were washed out post incubation. Osmotic shocks were performed while the slide was in the microscope by adding 24~$\mu$l of the shocking solution to one end and immediately bringing a piece of tissue paper to the other, resulting in flow and exchange of media. The flow duration was approximately 10-15~s and the local flow rate that the attached bacterium experienced was 0.68~$\mu$l/min \cite{Buda2016}. Tunnel was sealed after the shock to prevent evaporation and thus potential further increase in osmolarity throughout the course of the experiment.
For DDM experiments an $\sim$400~$\mu$m deep flat capillary (Vitrocom) glass sample was filled with $\sim$150~$\mu$l of bacterial suspensions immediately after upshock, and subsequently sealed to prevent evaporation during the experiment.

\subsection{Microscopy and data collection}
For BFM experiments backfocal plane interferometry \cite{Denk1990,Svoboda1993} was performed using a custom built microscope and a 855~nm laser (Blue Sky Research, USA) which formed  a weak optical trap. The rotating bead attached to a flagellar stub was brought into the focus of the laser and the back-focal plane of the condenser was imaged onto a position sensitive detector (PSD Model 2931, New Focus, USA). The voltage signal from the PSD was passed through an analog anti-aliasing filter (low pass, Bessel type filter with a cut off frequency of 2.5~kHz, Krohn-Hite Corporation, USA) and sampled at 10 kHz (PCIE-6351 DAQ, National Instruments). 
For DDM experiments imaging within the glass capillary was performed at 100~$\mu$m away from the bottom of the capillary to avoid any interaction with the glass wall.
The imaging begun within $\sim$~2min post upshock, and consisted of a time-series of phase-contrast images (Nikon TE300 Eclipse fitted with a Nikon Plan Fluor 10×Ph1 objective, NA=0.3, Ph1 phase-contrast illumination plate). Imaging was performed at 100~Hz sampling rate (Mikrotron high-speed camera (MC 1362) and frame grabber (Inspecta 5, 1-Gb memory)) for $\sim$ 40~s per movie duration and using 512x512 pixels field of view. Both DDM and BFM experiments were performed at room temperature (21$\pm$1$^{\circ}$C).

\subsection{Data analysis} Data collected during BFM experiments was analyzed in the following way. X and Y signals (obtained from voltages from  the PSD) were passed through a moving window discrete Fourier transform, where the window moves point by point and the window size was 1.684~s. Time traces of single motor speed obtained in such a way were processed to calculate the CW Bias using Equation \ref{eq:bias}. CW Biases were calculated for a fixed interval sizes (30~s in Fig.~2A, 3~min in Fig.~2C and 5~min in Fig.2B), and using a 60~s long moving window in Fig.~2D and SI Appendix Fig.~7B-E. To calculate $\overline{\tau}_{CCW}$ we first take a mean of the $\tau_{CCW}$ intervals belonging to each individual BFM rotation trace. We then pool so obtained mean cell intervals into a distribution and obtained its mean (similar is true for $\overline{\tau}_{CW}$, but in opposite direction). We include only bound CCW and CW intervals (see Fig.~11). In DDM experiments, the mean swimming speed was calculated from DDM movies as an average over $\sim$10$^4$~cells/ml. Details of the image processing and data analysis were as before \cite{Wilson2011,Martinez2012,Martinez2014,SwarzLinek2016}. DDM allows the measurement of an advective speed, simultaneously, over a range of spatial-frequency q, where each q defines a length-scale L=2*pi/q (for more details, see \cite{Wilson2011, Martinez2012}). A mean swimming speed is extracted by averaging over q. Data plotted in Fig.~3D were obtained by averaging over the range 0.5~$\lesssim$~q~$\lesssim$~2.2 $\mu$m$^{-1}$, corresponding to a range in length-scale 3~$\lesssim$~L~$\lesssim$~13~$\mu$m.


\section{Acknowledgments}
This work was supported by Chancellor's Fellowship to TP. WCKP and VM were funded by an EPSRC Programme Grant EP/J007404/1 and an ERC advanced grant (AdG 340877 PHYSAPS). We thank all of the members of the Pilizota lab, Poon Lab, and Alex Morozov and Filippo Menolascina for useful discussions and support.

\newpage
\section{SUPPORTING INFORMATION}
\subsection{Growth Media Osmolarities}
Growth media osmolalities were measured with an osmometer (Micro-Digital Osmometer MOD200 Plus, Camlab, Cambridge, UK) and are given in Table 1.
\subsection{Motility Media Viscosities}
Motility media viscosities were measured with a rheometer (TA Instruments AR2000, cone-plate geometry, 60 cm, 0.5$^{\circ}$) and are given in Table 2.

\subsection{Supporting Figures}
Supporting Figures to the main text are given with the captions.

\begin{table*}[p]
\centering
\caption{Media osmolalities}\label{Osmol/kg}
	\begin{tabular}{l c }
 Media 		& Osmolality (mOsmol/kg) 		\\ 
 \hline
    	VRB 				& 92			\\
    	VRB+40 mM sucrose				& 135 \\
        VRB+60 mM sucrose			& 157 \\
        VRB+100 mM sucrose		& 203			\\
    	VRB+200 mM sucrose	& 320			\\
    	VRB+400 mM sucrose			& 580				\\
        VRB+600 mM sucrose				& 877 \\
    \hline
	\end{tabular}
    \end{table*}

\begin{table*}[p]
\centering
\caption{Measured media viscosities (in agreement with listed in \cite{sucrose})}\label{Osmol}
	\begin{tabular}{l c }
		Media 		& Viscosity (Pa s)		\\ \hline
    	Motility Media 				& 0.00112 $\pm$ 0.00001	\\
    	Motility Media + 300 mM sucrose				& 0.00146 $\pm$ 0.00001  \\
        Motility Media + 600 mM sucrose			& 0.00192 $\pm$ 0.00002 \\
        Motility Meida + 900 mM	sucrose	& 0.00290 $\pm$ 0.00007			\\
    	Motility Media + 1850 mM sucrose	& 0.0172 $\pm$ 0.0001				\\
    \hline
	\end{tabular}
    \end{table*}

\begin{figure*}[p]
\centerline{\includegraphics[width=1\linewidth]{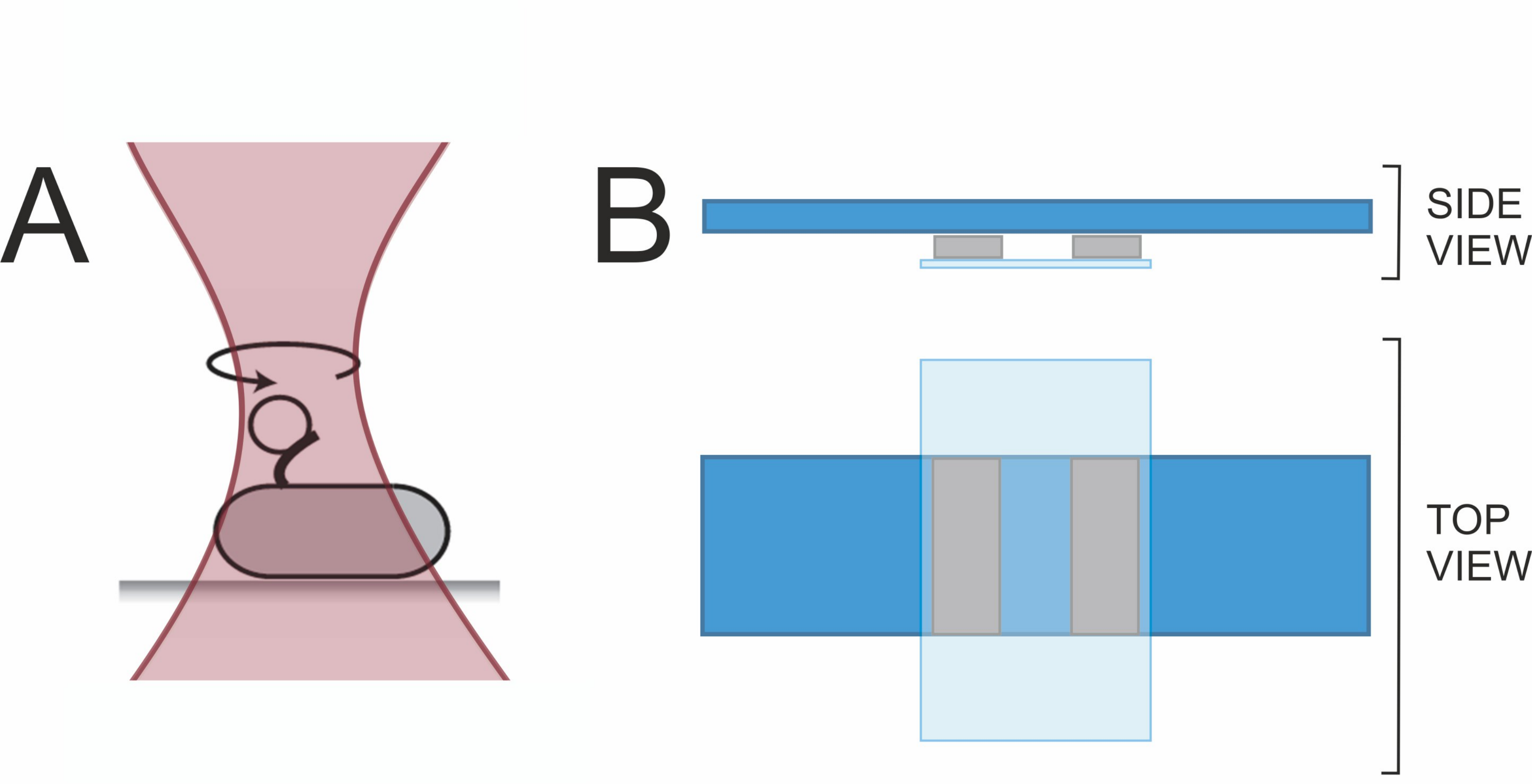}}
\caption{Schematic representation of the bead assay and tunnel slide used for measurements. (A) Genetically modified bacterial flagellar filament (FliC$^{Sticky}$ \cite{Fahrner1995}) stub was attached to a latex bead of 0.5~$\mu$m diameter. The bead was placed in a heavily attenuated optical trap and the bead rotation recorded via back focal plane interferometry (\textit{Methods}). (B) A tunnel slide, formed by a double sided sticky tape separating a microscope slide and a glass cover slip, was used to deliver osmotic shocks. The slide was sealed with vaseline upon the osmotic shock to prevent evaporation.}\label{cartoon}
\end{figure*}

\begin{figure*} [p]
\centerline{\includegraphics[width=1\linewidth]{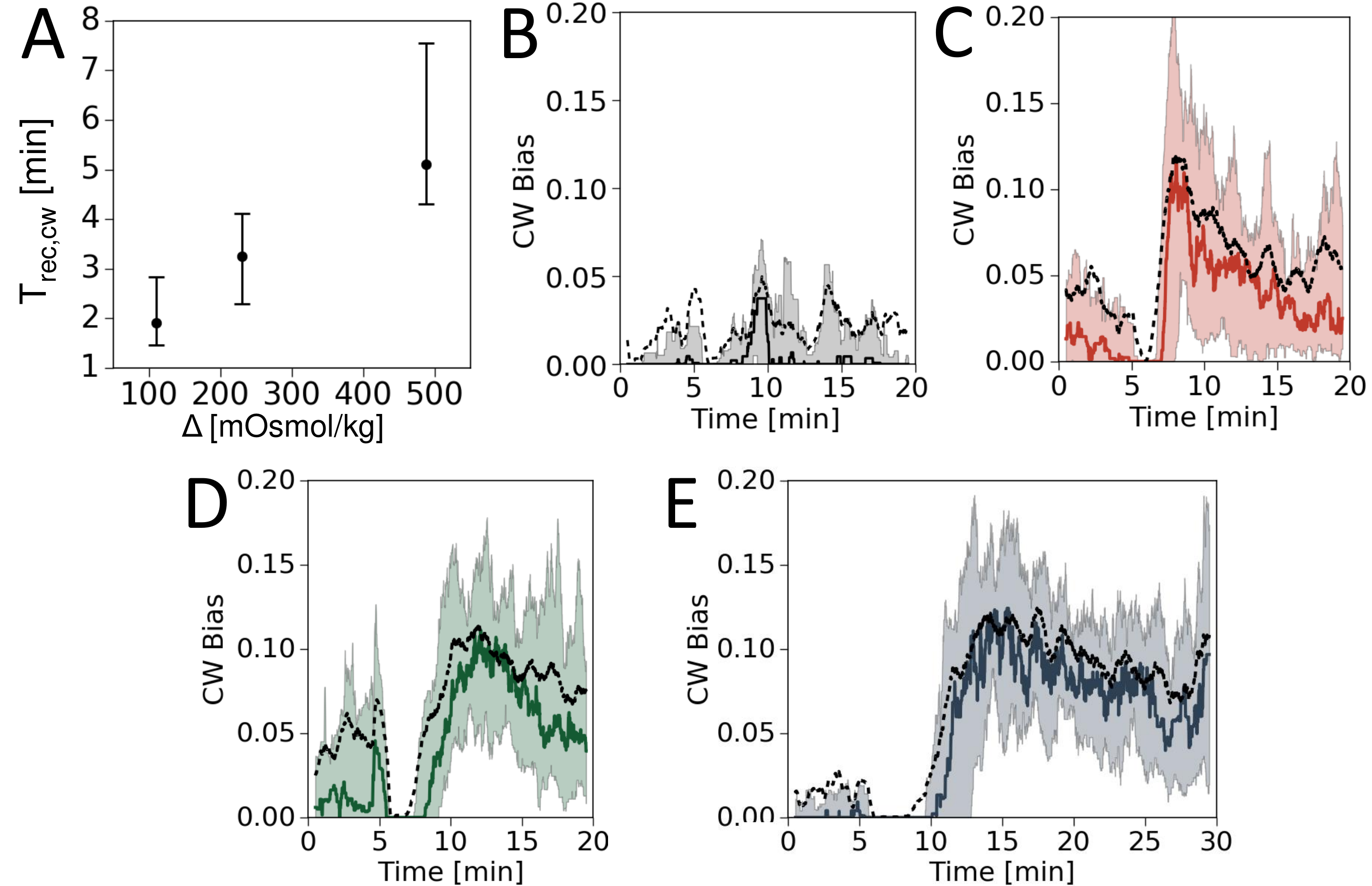}}
\caption{(A) Recovery Phase duration ($T_{rec,CW}$), calculated as the time needed for the post-shock median population bias to recover to the median VR Buffer bias (CW Bias=0.01, histogram in main text Fig. 2B). $T_{rec,CW}$ was calculated for each of the three shock conditions, whose biases are presented in panels C-D. The points represent times at which the post shock median reaches 0.01, relative to the time of shock delivery. The error bars give the populational variability in terms of recovery times for the 25th and the 75th percentile of population bias. (B) Buffer to buffer control flush CW Bias during the course of the experiment is given. Biases were calculated on a moving window of 60~s length and the VR Buffer was flown into the tunnel slide in the same way and duration as for the osmotic shock experiments. 12 cells in total are presented, the solid black line is the median, dashed is the mean and the shading represents the area between the 25th and 75th percentile. 
(C) Same as B but for the 111~mOsmol/kg osmotic upshift and the plot represents the median of the 22 single cell bias traces (thick red line), the mean (dashed black line) and the red shading is the area between the 25th and the 75th percentile. (D) and (E) are the same as C, only for the 230~mOsmol/kg (24 cells) and the 488~mOsmol/kg (23 cells) osmotic upshifts, respectively. 
}\label{AllBias}
\end{figure*}

\begin{figure*} [p]
\centerline{\includegraphics[width=0.7\linewidth]{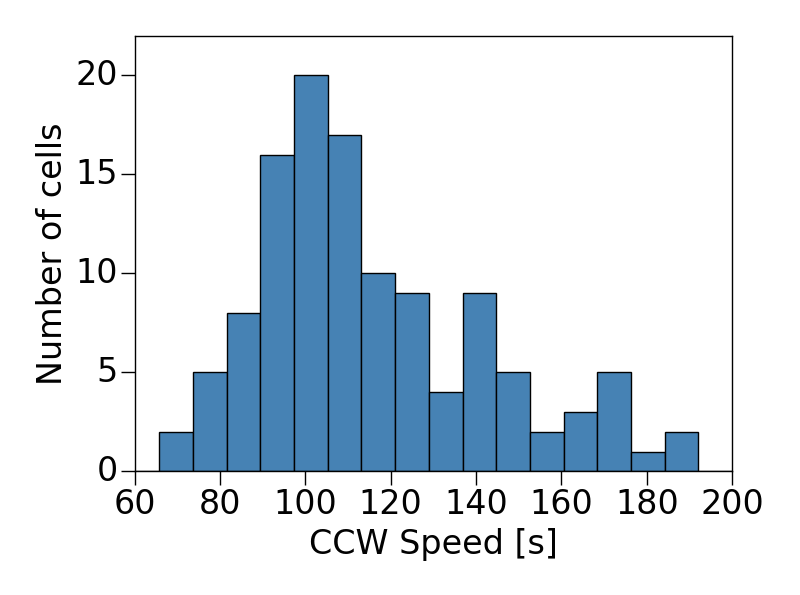}}
\caption{Histogram of single-motor speeds for \textit{E. coli} KAF84 in the Volume Recovery Buffer (VRB). Each motor comes from a different single cell. Each speed is computed by taking the first 15~s of various recordings made in VRB, selecting the CCW speed from the trace and averaging it. The histogram contains 118 cells and the mean value of the distribution is 116$\pm$2~Hz.}\label{AverageSpeed}
\end{figure*}

\begin{figure*} [p]
\centerline{\includegraphics[width=1\linewidth]{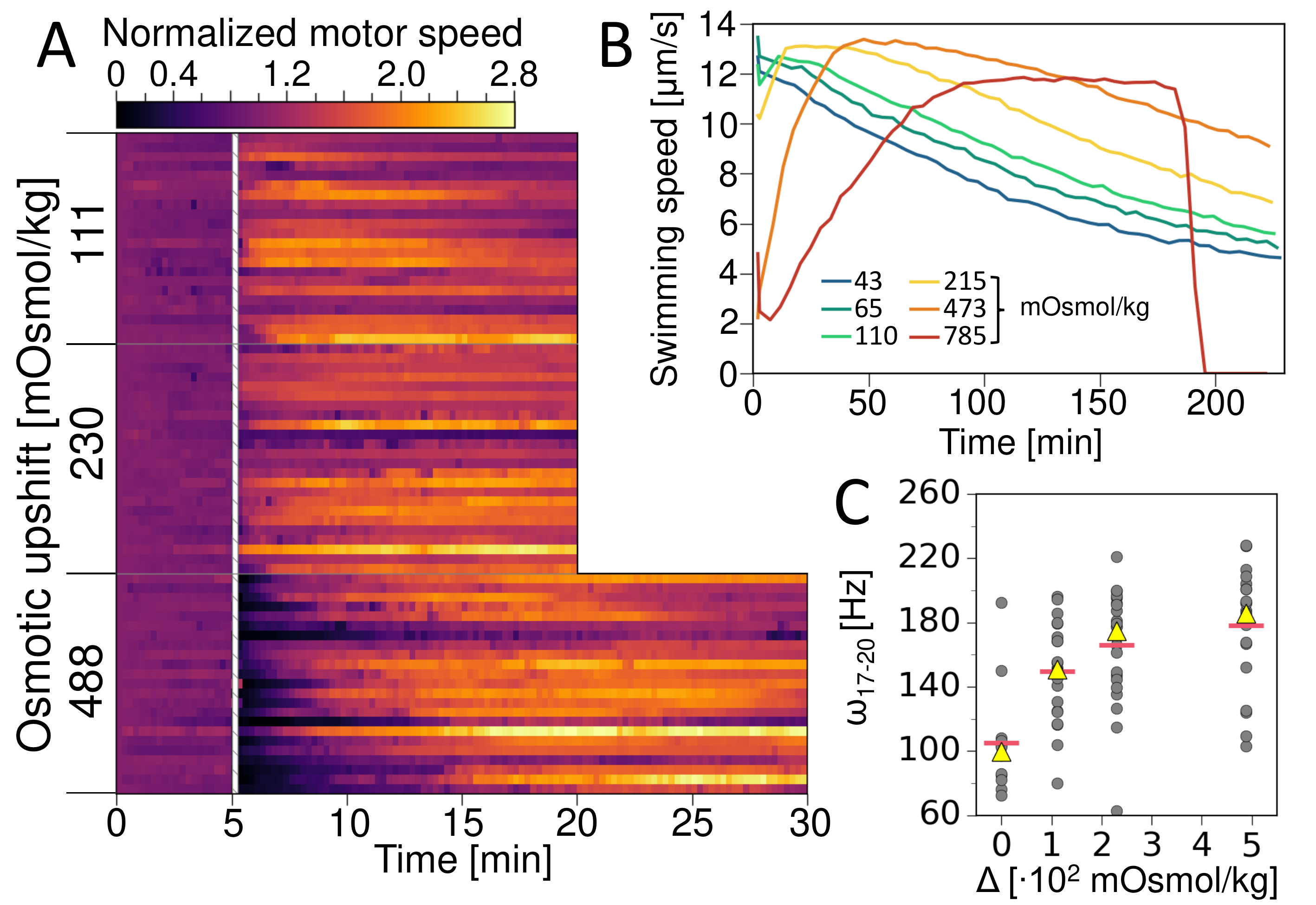}}
\caption{(A) Version of the main Fig.~3 where the motor and swimming speeds have been corrected for sucrose viscosity, under the assumption that no additional stators are incorporated at this increased load (see \textit{Discussion} in the main text). Viscosity correction for 100, 200 and 400~mM sucrose used are a multiplicative factor of 1.057, 1.195 and 1.444, respectively (measured and listed in \cite{sucrose}). Each line represents a single motor speed trace binned into 15~s intervals where the color represents the height of the bin, normalized to the first bin. Results are grouped by upshock magnitude, as indicated on the left hand edge. The white hatched column represents the point where an osmotic shock was administered by exchanging VR Buffer for VR Buffer + sucrose and the flow lasted for 10-15~s. There are 22, 24 and 23 cells for the 111, 230 and 488~mOsmol/kg condition, respectively. The color map scale is given at the top of the figure.
(B) DDM measurement of swimming speeds following an osmotic shock. Cells were shocked in microfuge tubes and brought into a microscope within 2~min. The legend shows shock magnitudes and the mean speed is the average of swimming speeds obtained for each time point in a range of different length scales (\textit{Methods}). The systematic error of our measurements is then calculated as the standard deviation of the mean, and falls within $\sim$5\% of the mean value (here not plotted for clarity). The traces shown have been viscosity corrected as well, as free-swimming cells with the motor operating in liner-torque regime do not increase the swimming speeds with viscosities we used in our experiments \cite{Martinez2014}.
(C) Viscosity corrected single motors speeds calculated as 3~min averages corresponding to a section between t=17 and t=20~min in A. The 0~mOsmol/kg condition is sampled from a set of buffer to buffer control flushes which were at least 20 minutes long (12 out of 18 control flushes). Red horizontal bars represent mean and yellow triangles median values.}\label{S4}
\end{figure*}

\begin{figure*} [p]
\centerline{\includegraphics[width=0.95\linewidth]{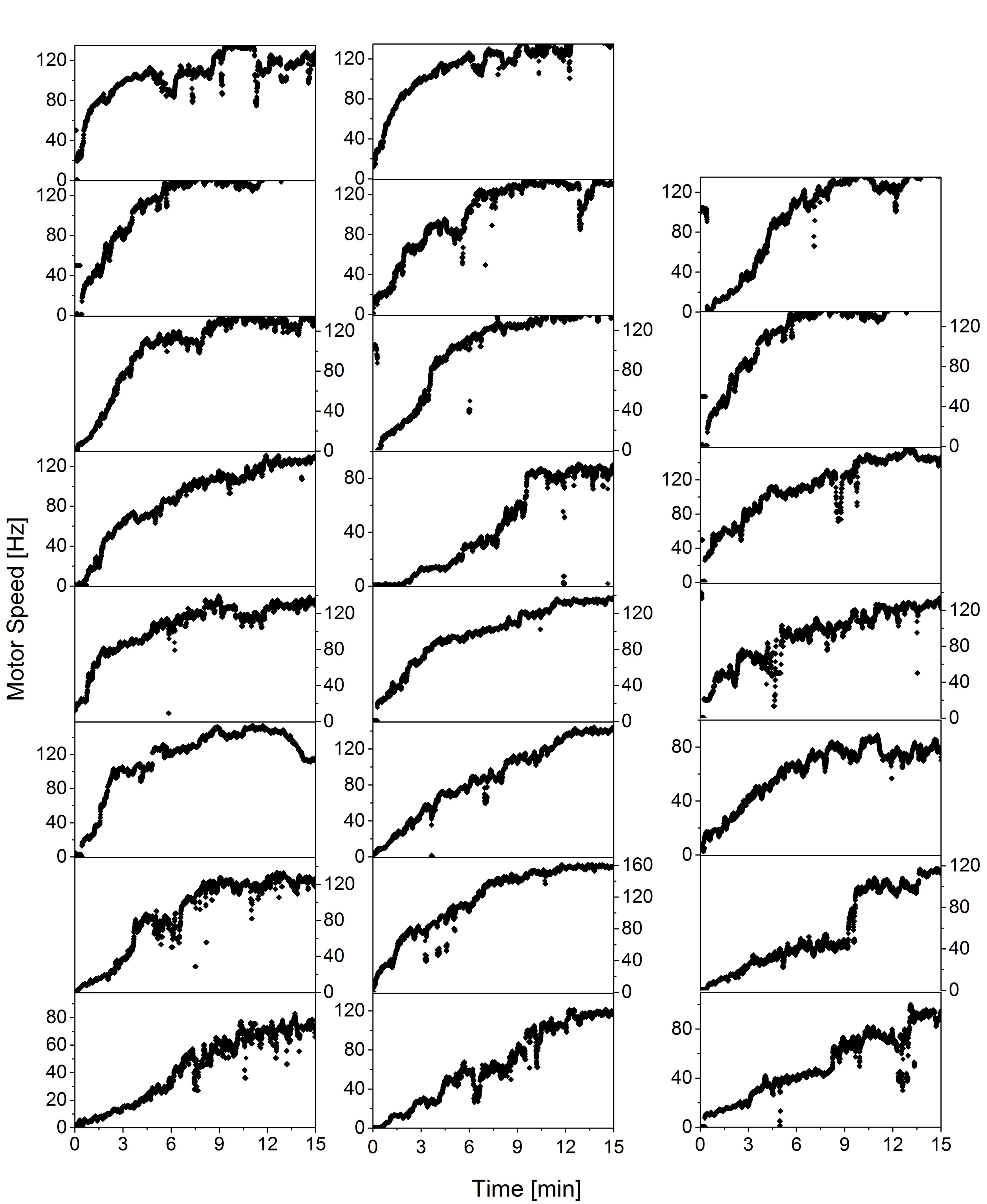}}
\caption{Traces of motor rotation after addition of 400~mM sucrose (all cells given in Fig.~2A and Fig.~3A of the main text are given). We do not plot the initial motor speed, and start at the point of speed drop due to an osmotic shock. Motor speeds were obtained as described in \textit{Methods}.}\label{PMFRec}
\end{figure*}

\begin{figure*} [p]
\centerline{\includegraphics[width=0.7\linewidth]{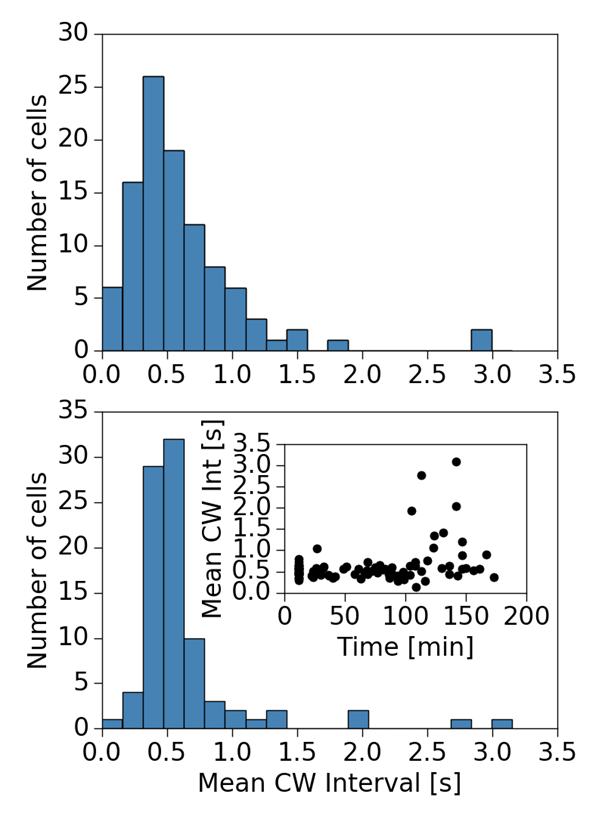}}
\caption{Top: Mean CW interval distribution ($\overline{{\tau}}_{CW}$) for cells in Volume Recovery Buffer. Motor rotation of 102 cells was sampled for 5~min and intervals counted as illustrated in SI Appendix Fig.~11. Bottom: Mean CW interval distribution for motors where the sampling interval (3~min) begins 12 or more minutes after an osmotic shock with 200~mM sucrose in VRB (+230~mOsmol/kg). The inset contains data from the histogram plotted against the start time of the sampling interval and t=0 corresponds to the osmotic shock.}\label{AverageSpeed2}
\end{figure*}

\begin{figure*} [p]
\centerline{\includegraphics[width=0.8\linewidth]{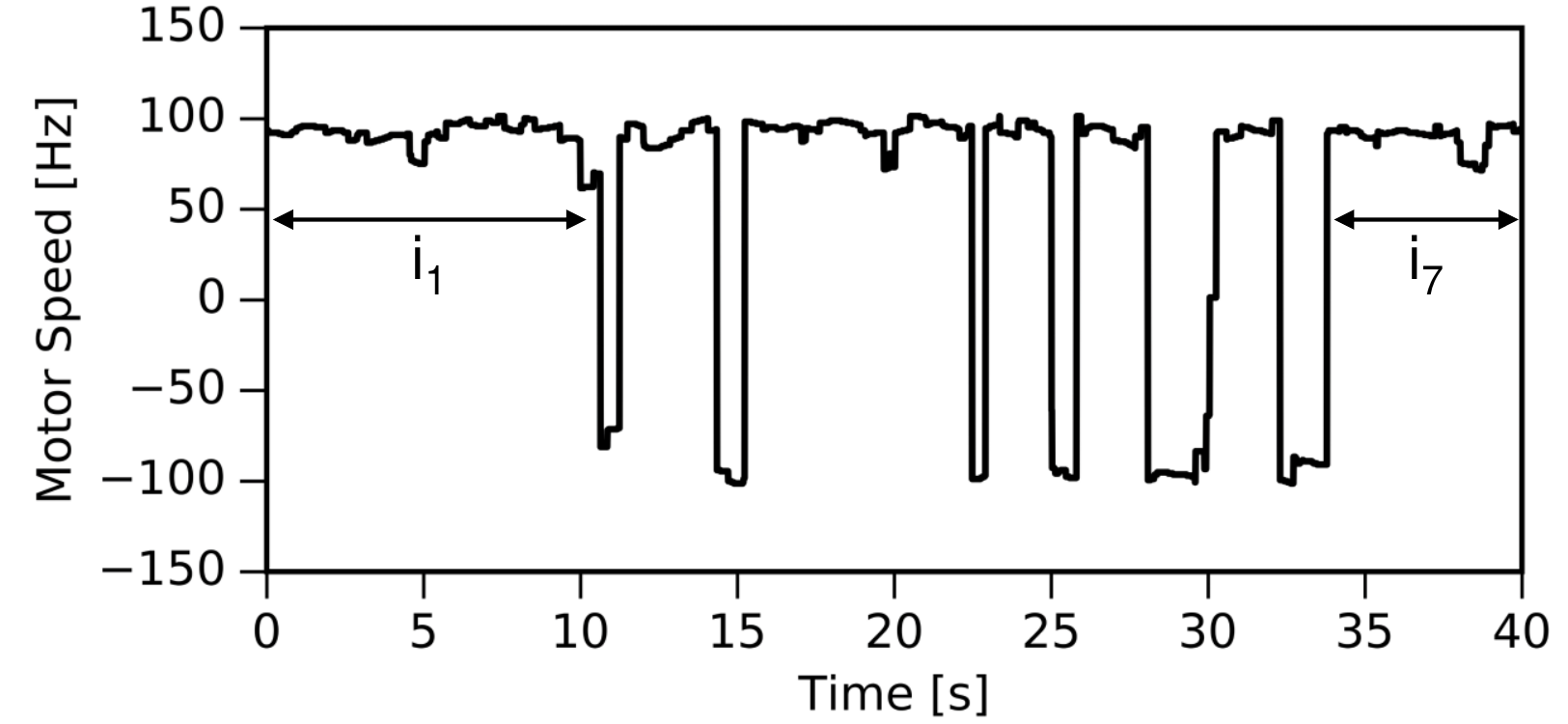}}
\caption{A portion of a single-motor speed trace illustrating the counting of interval lengths within a given sampling window. Positive speeds correspond to CCW rotation and negative to CW rotation. Intervals $i_1$ and $i_7$ are not included when calculating the mean CCW interval due to uncertainty in their beginning or end. Only intervals bounded by two CW intervals are counted.
Similar holds true when calculating mean CW interval, only those bounded by two CCW intervals were included.}\label{FigS7}
\end{figure*}

\FloatBarrier
\bibliographystyle{apalike}
\bibliography{sample}

\end{document}